\documentclass{chi-ext2}
\pdfoutput=1

\title{Does Anonymity Increase the Chance to Get Feedback?}

\numberofauthors{2}
\author{
  \alignauthor{
  	\textbf{Malte Paskuda}\\
  	\affaddr{ICD, HETIC, Tech-CICO, Troyes University of Technology, UMR 6281, CNRS}\\
  	\affaddr{12 rue Marie Curie - CS 42060}\\
  	\affaddr{10004 Troyes cedex, France }\\
  	\email{malte.paskuda@utt.fr}
  }\alignauthor{
  	  }
  \vfil
  \alignauthor{
  	\textbf{Myriam Lewkowicz}\\
  	\affaddr{ICD, HETIC, Tech-CICO, Troyes University of Technology, UMR 6281, CNRS}\\
  	\affaddr{12 rue Marie Curie - CS 42060}\\
  	\affaddr{10004 Troyes cedex, France }\\
  	\email{myriam.lewkowicz@utt.fr}
  }\alignauthor{
  	  }
}
\def\plaintitle{CHI LaTeX Extended Abstracts Template}
\def\plainauthor{Luis A. Leiva}
\def\plainkeywords{Anonymity, Online Participation, Youtube.}
\def\plaingeneralterms{Documentation, Standardization}

\hypersetup{
  pdftitle={\plaintitle},
  pdfauthor={\plainauthor},  
  pdfkeywords={\plainkeywords},
  pdfsubject={\plaingeneralterms},
}
\usepackage{graphicx}   
\usepackage{balance}    
\usepackage{bibspacing} 
\usepackage{tikz}
\usepackage{gincltex}
\usepackage{pgfplots}
\pgfplotsset{width=5cm}
\pgfkeys{
     /pgf/number format/precision=2, 
    /pgf/number format/fixed zerofill=false,
    /pgf/number format/fixed
}
\usepackage{subfigure}
\begin{document}
\maketitle
\section{Abstract}
To generate a hypothesis about the effects of anonymity on user participation in online communities, comments on Youtube were analysed for effects of the change from allowing pseudonyms to Google+ with its real name policy. Small differences were detected, leading to the hypothesis that the option to remain anonymous leads to a less active environment for getting feedback, with less polite and less rude comments on the expense of neutral ones.

.


\section{Introduction}
Shall people be anonymous on the internet? What effect has being anonymous in a group? While working on the concept for AAL TOPIC\footnote{\url{http://www.topic-aal.eu/}}, an online platform for social support among informal caregivers, the issue of allowing anonymous comments in the discussions or in general on the platform arose. It was evoked in particular by informal caregivers taking care of a person suffering from the Alzheimer's Disease, since this disease is related to behavioural disorders that can lead to embarrassing situations for the caregivers (and the patient). If the goal is to have a healthy community that engages in experience sharing and helping each other, would it be better to ask users to use their civil identity or do they gather confidence by being anonymous or pseudonymous? Or is it better to allow both situations, depending on the type of discussions?

With theories and literature implying different outcomes (\cite{anonIntergroup}, \cite{SidePortraits}), it became clear that this question is not answered easily. We decided to look at comments in Youtube, especially because on Nov. 2013, Google integrated Youtube's comment system into Google+. Before the change, users were free to chose a name, but after the change, users were forced to use their full civil identity (later, pseudonyms were allowed, but the character of the platform changed) \footnote{Causing several Youtubers to forbid comments, see \url{http://goo.gl/wkkbBy}}. Thus, we can find videos with comments made by users with pseudonym only, and newer videos where commenters often use their full name, while being connected by Google+ to their friends and identity.

This situation gives us the option to compare: 
\begin{enumerate}
\item Comments from before and after the change.
\item Comments from before the change by users with and without a Google+ account now.
\end{enumerate}

This is a first step in a bigger effort to analyse the impact of anonymity and to find recommendations for community builders. The hypotheses generated here are planned to be tested in other studies and in an experiment as part of the TOPIC project. In the following section, we present the related work. Afterwards we show a simplified model built from the literature, describing the relation between anonymity and participation. The section after describes the data gathering and the findings while following the model. Limitations are mentioned and a conclusion is made.

\section{Related Work}

Research work on the effect of anonymity already exists. A fundamental theory is the Deindividuation Theory, describing how a member in a group looses his self-awareness, thus loosing his social conscience \cite{anonIntergroup}, which leads to less polite discussions. A second theory, the social identity model of deindividuation effects (SIDE), regards anonymous group behaviour more positive. It suggests that members try to do what is good for the group, because members identify with the norms of the group \cite{SidePortraits}, with anonymity helping that process. The user has a better user experience by feeling more connected.

An important practical work is described by Kilner et al in \cite{CoPAnonymity}. An online forum for soldiers gradually changed its account model from anonymity with pseudonyms to full civil identity. Kilner et al. analysed the different stages and found that removing anonymity options led to fewer antisocial comments and fewer comments in total. This work heavily influenced our analysis in selecting possible hypotheses.

In the area of Behavioral Science, experiments (like \cite{anonymityDisinhibition}) tried to find effects of anonymity. Research on the link between politeness, civility and anonymity from a political angle (\cite{santana2012civility}) exists as well.

There is also a lot of literature describing factors influencing participation. Anonymity is there seldom a main focus, but it gets mentioned. An example for that is a main thread in the literature: The \textbf{Common Identity and Bond Theory} being used by Kraut et al as described in \cite{commonBond}. The theory sees two types of connection between the members of a community - Identity and Bond - influenced by different factors, \emph{Social Categorization}, \emph{Interdependence} and \emph{Intergroup Comparisons} for Identity and \emph{Social Interaction}, \emph{Personal Information} and \emph{Personal Attraction through Similarity} for Bond. Some of them can be linked to anonymity, SIDE theory does that explicitly with \emph{Personal Attraction through Similarity}, as described in the introduction.

\section{Model}
\begin{figure}
    \centering
    \centerline{\includegraphics[scale=0.5]{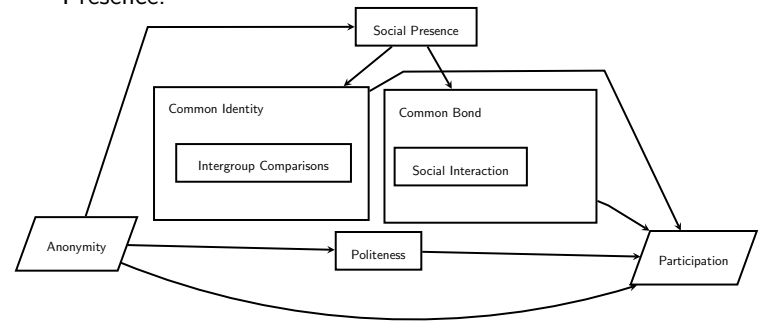}}
    \caption{simplified model showing the interaction of anonymity and participation}
    \label{fig:model}
\end{figure}

As we are interested in the influence of anonymity on online participation for social support and community building, we developed a model of what influences participation. We started with Kraut's use of the Common Identity and Common Bond model, and other related work. Then, factors influenced by anonymity were collected. We ended with a model showing which factors that might influence participation are influenced by anonymity. But many factors were hard to apply on a textual corpus. We then simplified the model, keeping only the factors for which we were able to find markers we can measure in text (see \ref{fig:model}). 

This model illustrates that a big part of the literature is assuming that anonymity influences politeness (\cite{offensiveInternet}). Politeness was found to influence participation; For instance \cite{bbcNegative} showed that impolite comments provoke other comments. Anonymity is indirectly connected with Intergroup Comparisons and Social Interaction via Social Presence; \cite{socialPresence} describe that factors linked to Common Bond and Common Identity profit from Social Presence, with \cite{anonPresence} showing that anonymity influences Social Presence.

\section{Data Collection}
24 videos were identified that had several comments and were related to informal caregivers or Alzheimer. The average publishing date of the comments was Monday, December 6, 2010. The 3773 comments were downloaded with Youtubes API (using modified scripts of the TubeKit parser\footnote{\url{http://tubekit.org/}}), as well as the profile information of the 3087 users, revealing whether the account was linked with Google+ or not. Youtubes API does not show when users linked their Youtube-Account to Google+, we can only see which commenters are still not using Google+. However, it is complicated to use Youtube while logged in without going through the Google+ boarding. Consequently, no comment made after the change to Google+ was from a user without Google+. The other way around existed, there were comments from people having only a Google+ account and no Youtube profile, but all were discarded for being formal sharing announcements.

The comments were then analysed for markers that showed: 
\begin{enumerate}
\item \textbf{Politeness}. To measure those factors in text we searched for markers that show how polite a message is. After dismissing some models as too complicated to use manually (\cite{politenessMarkers}) or not accessible enough (\cite{computationalPoliteness}), it was decided to use an algorithmical approach (Bayes' algorithm). 
\item \textbf{Intergroup Comparisons}. We searched for the words "we/us/our/them", that grammatically show that a group of people is mentioned. In the model, the use of intergroup comparisons is influenced by anonymity through \textbf{social awareness}.
\item \textbf{Personal Interaction}. We looked at the reply count given by Youtubes API. In the model, this is influenced by anonymity through \textbf{social awareness}. The amount of replies made has to be fetched from the comment data by searching for the @-character, this metric worked only before the change to Google+.
\end{enumerate}

The use of the Bayes' algorithm was thereby the most complicated step. 300 comments were marked manually as either polite, neutral or rude. Then the algorithm classified all remaining comments. The classification of 100 comments was used to calculate an estimated accuracy. The accuracy of the used algorithm was 80\%.

\section{Findings}
\subsection{The change}\label{sc:change}
Comparing comments from before and after the change, there is a  difference.
\subsubsection{Politeness}
 After the change, we find slightly more polite and rude comments, significant by a \(\chi^2\)-test with \(p < 0.01\).

\begin{center}
\begin{tabular}{| l | c | c | c | }
    \hline
     & Polite & Neutral & Rude \\
    \hline
    Before & 133 (3\%) & 2838 (92\%)& 155 (5\%) \\
    After & 32 (5\%) & 534 (84\%) & 81 (11\%)\\
 \hline
\end{tabular}
\end{center}

\subsubsection{Intergroup Comparisons}
Most of the comments did not contain intergroup comparisons (we/us/our/them). After the change, the average use of those words was slightly higher, but a t-test showed the increase to be not significant.

\begin{table}
\caption{Amount of Comparisons}
\centering
\begin{tabular}{| l | c | c | c | c |}
\hline
\hline
Group & mean & sd & median & n \\
\hline
Before & 0.1628 & 0.5885 & 0 & 3126 \\
After & 0.2365 & 1.3417 & 0 & 647 \\
\hline
\end{tabular}
\label{table:comparisonsChange}
\end{table}

\subsubsection{Social Interaction}
There are two different metrics for social interaction in the data: replies made and replies gotten. The Youtube API only shows the amount of replies gotten. The difference when looking at the effect of the change is big, and significant by t-test with \(p < 0.01\) . After the change, with an average of 0.5 it seems like every second comment was answered, though the median of 0 shows this to be false. Instead some comments got many replies, while many other still got none.

\begin{table}
\caption{Change of replies}
\centering
\begin{tabular}{| l | c | c | c | c |}
\hline
\hline
Group & mean & sd & median & n \\
\hline
Before & 0.0067 & 0.1171 & 0 & 3126 \\
After & 0.4791 & 2.3598 & 0 & 647 \\
\hline
\end{tabular}
\label{table:repliesChange}
\end{table}

\subsection{Pseudonymous vs Google+ Users before the change}

We just saw that the change in the environment had an influence on the comments. But that does not prove that the change in the degree of anonymity is the cause of that change, as other factors changed as well. A difference in the comments between users who adopted Google+ and those who did not would be a clearer signal, but the difference was small.

\subsubsection{Politeness}
There was no difference in the politeness rating, confirmed by a \(\chi^2\)-test resulting in \(p = 0.8424\).

\begin{center}
\begin{tabular}{| l | c | c | c | }
    \hline
     & Polite & Neutral & Rude \\
    \hline
    G+ & 96 (4\%) & 2058 (91\%) & 112 (5\%) \\
    pseudonym & 36 (4\%) & 730 (91\%) & 36 (4\%) \\
 \hline
\end{tabular}
\end{center}

\subsubsection{Intergroup Comparisons}
Intergroup Comparisons made were also on the same level.

\begin{center}
\begin{tabular}{| l | c | c | }
    \hline
     & Comments With Comparisons \\
    \hline
    G+ & 253 (10\%) \\
    pseudonym & 83 (10\%) \\
 \hline
\end{tabular}
\end{center}

\subsubsection{Social Interaction}
The only visible difference is here. According to the API, no pseudonymous user got any reply. They made however the same relative amount of replies. The lack of responses could explain why the users stopped being active (see \cite{feedbackWiki}). This observation could be a bug in the API, but is not totally unlikely given low amount of replies. Comments were often directed at the creator of the video, not at other commenters. Sadly the identification whether a comment was a reply or not was not reliable. That data is not coming from the API but from searching for an "@" sign, a praxis used before the change to reference another user.

\begin{center}
\begin{tabular}{| l | c | c | }
    \hline
     & Avg Replies Gotten & Avg of being a Reply \\
    \hline
    G+ & 0.01  & 0.085 \\
    pseudonym & 0 & 0.081 \\
 \hline
\end{tabular}
\end{center}

\section{Limitations}
It is possible that the markers that were measured are influenced by other factors, and that anonymity did not play a significant role. Youtube changed its interface, the spam control and the ranking of comments, from a timeline system showing the newest comments first to an opaque ranking system. External cultural factors could also influence the comments. Thus a different selection of videos could show other results. Another limitation is the bayes algorithm used to qualify politeness. The initial supervised learning process depends on the researcher entering the input. The observed 80\% accuracy is subject to the same limitation, as the algorithmic politeness rating was compared with the subjective correct rating.

\section{Conclusion and Further work}
Given the limitations of this study, the results are rather hypotheses for further work. There are two: (1) When commenters are anonymous, it leads to less polite and less rude comments. (2) When commenters are anonymous, it leads to less interaction.

The first hypothesis is especially surprising, as it stands in contrast to what was found by Kilner et al in \cite{CoPAnonymity}. It is further interesting to see that there was no difference observed between the commenters using Google+ now and those who chose to stay pseudonymous, or to abandon Youtube after the change, apart from the reply count. The expectation when looking at that data was to see a difference caused by a different mentality of those accepting Google+ and those who did not. Further research is needed to work around the limitations of this analysis. A new study will look at a truly mixed environment, where anonymous members and those showing their civil identity are members at the same time (Wikipedia.org for instance). Another study will look at environments that use different identity models but are related, like discussion boards for similar topics (for example 4chans /g/ and Hacker News).

\section{Acknowledgments}
This work has been supported by European Union, ANR and national solidarity fund for autonomy through AAL program (project AALI 2012-TOPIC).

\balance
\bibliographystyle{acm-sigchi}
\bibliography{main}

\begin{thebibliography}{10}

\bibitem{bbcNegative}
Chmiel, A., Sobkowicz, P., Sienkiewicz, J., Paltoglou, G., Buckley, K.,
  Thelwall, M., and Hołyst, J.~A.
\newblock Negative emotions boost user activity at \{BBC\} forum.
\newblock {\em Physica A: Statistical Mechanics and its Applications 390}, 16
  (2011), 2936 -- 2944.

\bibitem{SidePortraits}
Cress, U.
\newblock Why member portraits can undermine participation.
\newblock In {\em Proceedings of the 2005 Conference on Computer Support for
  Collaborative Learning: Learning 2005: The Next 10 Years!}, CSCL '05,
  International Society of the Learning Sciences (2005), 86--90.

\bibitem{computationalPoliteness}
Danescu{-}Niculescu{-}Mizil, C., Sudhof, M., Jurafsky, D., Leskovec, J., and
  Potts, C.
\newblock A computational approach to politeness with application to social
  factors.
\newblock {\em CoRR abs/1306.6078\/} (2013).

\bibitem{socialPresence}
Farzan, R., Dabbish, L.~A., Kraut, R.~E., and Postmes, T.
\newblock Increasing commitment to online communities by designing for social
  presence.
\newblock In {\em Proceedings of the ACM 2011 Conference on Computer Supported
  Cooperative Work}, CSCW '11, ACM (New York, NY, USA, 2011), 321--330.

\bibitem{politenessMarkers}
House, J., and Kasper, G.
\newblock Politeness markers in english and german, in conversational routine:
  Explorations in standardized communication patterns and prepatterned, 1981.

\bibitem{CoPAnonymity}
Kilner, P.~G., and Hoadley, C.~M.
\newblock Anonymity options and professional participation in an online
  community of practice.
\newblock In {\em Proceedings of th 2005 Conference on Computer Support for
  Collaborative Learning: Learning 2005: The Next 10 Years!}, CSCL '05,
  International Society of the Learning Sciences (2005), 272--280.

\bibitem{anonymityDisinhibition}
Lapidot-Lefler, N., and Barak, A.
\newblock Effects of anonymity, invisibility, and lack of eye-contact on toxic
  online disinhibition.
\newblock {\em Computers in Human Behavior 28}, 2 (2012), 434 -- 443.

\bibitem{offensiveInternet}
Levmore, S., Levmore, S., and Nussbaum, M.
\newblock {\em THE OFFENSIVE INTERNET}.
\newblock Harvard University Press, 2010.

\bibitem{anonIntergroup}
Postmes, T., Spears, R., and Lea, M.
\newblock Intergroup differentiation in computer-mediated communication:
  Effects of depersonalization.
\newblock {\em Group Dynamics: Theory, Research, and Practice 6}, 1 (2002), 3.

\bibitem{commonBond}
Ren, Y., Kraut, R., and Kiesler, S.
\newblock Applying common identity and bond theory to design of online
  communities.
\newblock {\em Organization studies 28}, 3 (2007), 377--408.

\bibitem{santana2012civility}
Santana, A.
\newblock Civility, anonymity and the breakdown of a new public sphere.

\bibitem{anonPresence}
Tu, C.-H.
\newblock The relationship between social presence and online privacy.
\newblock {\em The Internet and Higher Education 5}, 4 (2002), 293 -- 318.

\bibitem{feedbackWiki}
Zhu, H., Zhang, A., He, J., Kraut, R.~E., and Kittur, A.
\newblock Effects of peer feedback on contribution: A field experiment in
  wikipedia.
\newblock In {\em Proceedings of the SIGCHI Conference on Human Factors in
  Computing Systems}, CHI '13, ACM (New York, NY, USA, 2013), 2253--2262.

\end{thebibliography}

\end{document}